\documentclass[aps,prl,twocolumn,english,showpacs,superscriptaddress,amssymb,amsfonts]{revtex4}
\usepackage[T1]{fontenc}
\usepackage[latin9]{inputenc}
\usepackage{amsmath}
\usepackage{dsfont}
\usepackage{amstext}
\usepackage{pdfsync}

\usepackage{tocvsec2}

\usepackage{amssymb}
\usepackage{amsbsy}
\usepackage{amsthm}
\usepackage{epsfig}
\usepackage{framed}
\usepackage{graphicx}
\usepackage{bbm}
\usepackage{hyperref}
\usepackage{color}
\usepackage{multirow}

\newcommand{\bst}{{\mathcal{T}}}

\newcommand{\ie}{{\emph{i.e.~}}}
\makeatletter

\newcommand{\Rmnum}[1]{\expandafter\@slowromancap\romannumeral #1@}
\makeatother
\newcommand{\imth}{\hspace{1pt}\mathrm{i}\hspace{1pt}}

\newcommand{\eg}{{\emph{e.g.~}}}

\newcommand{\bea}{\begin{eqnarray}}
\newcommand{\eea}{\end{eqnarray}}
\newcommand{\bpm}{\begin{pmatrix}}
\newcommand{\epm}{\end{pmatrix}}
\newcommand{\bal}{\begin{aligned}}
\newcommand{\eal}{\end{aligned}}

\newcommand{\dket}[1]{|{#1}\rangle}

\newtheorem{theorem}{Theorem}

\makeatother

\usepackage{babel}
\begin{document}
\title{Symmetry-enforced quantum spin Hall insulators in $\pi$-flux models}

\author{Jiaxin Wu}
\author{Tin-Lun Ho}
\author{Yuan-Ming Lu}
\affiliation{Department of Physics, The Ohio State University, Columbus, OH 43210, USA}

\begin{abstract}
We prove a Lieb-Schultz-Mattis theorem for the quantum spin Hall effect (QSHE) in two-dimensional $\pi$-flux models. In the presence of time reversal, $U(1)$ charge conservation and magnetic translation (with $\pi$-flux per unit cell) symmetries, if a generic interacting Hamiltonian has a unique gapped symmetric ground state at half filling (i.e. an odd number of electrons per unit cell), it can only be a QSH insulator. In other words, a trivial Mott insulator is forbidden by symmetries at half filling. We further show that such a symmetry-enforced QSHE can be realized in cold atoms, by shaking an optical lattice and applying a time-dependent Zeeman field.
\end{abstract}

\pacs{}

\maketitle




\paragraph{Introduction---} As the first theoretically predicted topological insulator\cite{Hasan2010,Qi2011}, quantum spin Hall effect (QSHE) in two spatial dimensions has drawn much attention due to its physical novelty and potential applications\cite{Konig2008,Qi2010,Maciejko2011}. In spite of numerous theoretical efforts, so far its experimental realization has been largely restricted to semiconductors, where a ``band inversion'' driven by strong spin-orbit couplings can turn a trivial band insulator into a nontrivial QSH insulator. It is also unclear how this band-inversion mechanism for QSHE can be extended to a generic interacting system.

One definitive feature of QSHE is that a $\pi$-flux excitation therein can carry a half-integer spin and form a Kramers doublet\cite{Ran2008a,Qi2008a}. Inspired by this property, we show that the magnetic translation symmetry can serve as a new mechanism for QSHE at half filling, i.e. with an odd number of electrons (and a $\pi$-flux) per unit cell. Specifically we prove a Lieb-Schultz-Mattis (LSM) type theorem, which forbids a trivial band insulator ground state: it dictates any short-range-entangled ground state that preserves all symmetries \emph{must} be a QSH insulator. Applicable to a generic system, this theorem sheds new light in the search of QSHE in strongly-interacting systems.

We demonstrate the power of this theorem in a simple model on square lattice. We further show that this model with magnetic translation symmetry can be realized by cold atoms in a shaking optical lattices, when a time-dependent Zeeman field is applied.

\paragraph{A LSM theorem for QSHE in $\pi$-flux models---}We consider a generic interacting (half-integer-spin) fermion system on any two-dimensional lattice with magnetic translation symmetry
\bea\label{mag trans sym}
\tilde T_1\tilde T_2\tilde T_1^{-1}\tilde T_2^{-1}=(-1)^{\hat F}
\eea
where $\tilde T_{1,2}$ are magnetic translations associated with Bravais lattice vectors $\vec a_{1,2}$, and $\hat F$ is the conserved total fermion number in the system. Magnetic translation symmetry (\ref{mag trans sym}) simply indicates a $\pi$ flux threaded through each unit cell on the 2d lattice. We also require time reversal symmetry $\bst$ satisfying
\bea\label{trs}
\bst^2=(-1)^{\hat F}
\eea
for fermions with half-integer spins. Our no-go theorem of Lieb-Schultz-Mattis (LSM) type states the following:

\begin{theorem}
\label{thm}
Consider a generic interacting (half-integer-spin) fermion system with $U(1)$ charge conservation, time reversal (\ref{trs}) and magnetic translation (\ref{mag trans sym}) symmetries. With $\bar\rho_f=$~odd fermions per unit cell, if there is a unique insulating ground state with no topological degeneracy that preserves all symmetries, it must be a QSH insulator.
\end{theorem}

This theorem has an important difference as compared to other LSM theorems proved earlier\cite{Oshikawa2000,Hastings2004,Hastings2005,Oshikawa2006,Parameswaran2013,Watanabe2015}, in the following sense. Usually a LSM theorem completely rules out the possibility of any short-range-entangled (SRE) symmetric ground state: \ie the ground state is either long-range entangled (gapless or intrinsic topological orders with ground state degeneracy), or spontaneously breaks the symmetry. In comparison, Theorem \ref{thm} allows the possibility of a nontrivial SRE symmetric ground state \ie a symmetry-protected topological (SPT) phase\cite{Chen2013,Senthil2014}. These SPT phases cannot be connected to a trivial product state (\eg a trivial Mott insulator) smoothly without either breaking symmetries or closing the bulk gap (via a phase transition). More interestingly they exhibits symmetry-protected gapless edge excitations, such as the helical edge modes in a QSH insulator here. Theorem \ref{thm} can be generalized to other symmetries, hence suggesting a new mechanism to realize SPT phases of matters.

Before giving a formal proof, we first present an intuitive argument for Theorem \ref{thm}. In a system with particle number conservation, time reversal (\ref{trs}) and translational symmetries, any SRE ground state must have a Kramers singlet (\ie an even number of fermions) per unit cell as proved in Ref.\cite{Watanabe2015}. However in our half-filled system with magnetic translations (\ref{mag trans sym}), there is only 1 fermion in addition to a background $\pi$-flux in each unit cell. In a trivial Mott insulator, a $\pi$-flux carries a trivial representation of time reversal symmetry, leading to 1 Kramers doublet in each unit cell, incompatible to a SRE symmetric ground state at half filling. On the other hand, it's well-known that an interacting character for a 2d QSH insulator is that each $\pi$-flux traps a Kramers doublet as protected by time reversal symmetry\cite{Ran2008a,Qi2008a}. As a result 1 fermion and 1 $\pi$-flux provides 2 Kramers doublets (hence a Kramers singlet) per unit cell, and therefore only a QSH insulator ground state is compatible with magnetic translation and time reversal symmetries at half filling.

\paragraph{Proof of The Theorem---}Below we provide a proof, combining a flux-insertion argument with an entanglement-spectrum argument\cite{Pollmann2010,Watanabe2015}. Without loss of generality we consider a square lattice on an infinite cylinder with a circumference $L_y=$~odd. We can always choose a Landau gauge for magnetic translation (\ref{mag trans sym})
\bea\label{landau gauge}
&\tilde T_y=T_y,~~~\tilde T_x=T_x\cdot(-1)^{\sum_{\bf r}y\hat n_{\bf r}},\\
&A_x({\bf r})=0,~~~A_y({\bf r})=x\pi.
\eea
where $T_{x,y}$ are pure translations, $A_{x,y}$ are vector potentials for fermion hopping phases, and $\hat n_{\bf r}$ is the fermion number on lattice site ${\bf r}=(x,y)$. Notice that magnetic translation $\tilde T_y$ is not well-defined on our cylinder with $L_y=$~\emph{odd} circumference, and hence absent in our system. However, as will be shown below, there is an emergent many-body symmetry for the unique SRE ground state in our system\footnote{Yuan-Ming Lu, Ying Ran and Masaki Oshikawa, to appear.}.

Imagine we adiabatically insert a flux of $L_y\Phi_0/2$ ($\Phi_0$ being the flux quantum) through the hole of the cylinder, during which there is always a finite excitation gap $\Delta_{min}>0$ between the unique SRE ground state and all excited states in the many-body system. In particular the flux insertion operator is given by
\bea
\mathcal{F}_y(\pi L_y)\equiv\mathcal{T}e^{-\imth\int_0^T\hat H[\phi_y(t)]\text{d}t}
\eea
where $\hat H[\phi(t)]$ is the time-dependent (always gapped) Hamiltonian during the adiabatic flux insertion process, with $\phi_y(0)=0$ (no flux) at $t=0$ and $\phi_y(T)=\pi L_y$ at $t=T$. Note that after the flux insertion, the vector potential along $\hat y$ direction is changed into
\bea
A_y^\prime({\bf r})=A_y({\bf r})+\pi L_y/L_y=(x+1)\pi=A_y({\bf r}+\hat x)
\eea
where $\vec A~(\vec A^\prime)$ is the vector potential before (after) the flux insertion. Clearly the whole lattice system is simply shifted by one lattice constant along $\hat x$ direction in the flux insertion process, and hence the unique SRE many-body ground state $|\Psi_0\rangle$ must be mapped to itself (up to a phase $\phi_0$) after the combined operation
\bea\label{new sym.}
\tilde{\mathcal{F}_y}\equiv T_x^{-1}\cdot\mathcal{F}_y(\pi L_y),~~~\dket{\Psi_1}\equiv\tilde{\mathcal{F}_y}|\Psi_0\rangle=e^{\imth\phi_0}|\Psi_0\rangle
\eea
Hence unitary operator $\tilde{\mathcal{F}_y}$ serves as an emergent many-body symmetry on the odd-circumference cylinder, which plays a crucial role in our proof.

Next we consider the Schmidt decomposition of unique SRE ground state $\dket{\Psi_0}$ across an entanglement cut along $\hat y$ direction located at $x_0-1<\bar x<x_0$ :
\bea
\dket{\Psi_0}=\sum_{\alpha}\lambda_{\bar x,\alpha}\dket{\alpha}_{\bar x,L}\dket{\alpha}_{\bar x,R}
\eea
where $\lambda_{\bar x,\alpha}$ are Schmidt weights. Note that each Schmidt state $\dket{\alpha}_{\bar x,L/R}$ is an eigenstate of fermion number operator ${\hat Q}_{\bf r}\equiv{\hat n}_{\bf r}-\bar\rho_f$
\bea
\sum_{x<\bar x}{\hat Q}_{\bf r}\dket{\alpha}_{\bar x,L}=Q_{\alpha,\bar x}\dket{\alpha}_{\bar x,L}
\eea
where $Q_{\alpha,\bar x}$ is the number fluctuation relative to average density $\bar\rho_f$ in the left region $x<\bar x$, which is well-defined in the thermodynamic limit\cite{Watanabe2015}. Under time reversal symmetry operation $\hat\bst$, Schmidt eigenstate $\dket{\alpha}_{\bar x,L}$ either transforms as a Kramers singlet or doublet depending on $\bst^2=\pm1$:
\bea
\hat\bst^2\dket{\alpha}_{\bar x,L}=(\bst^2)_{\alpha,\bar x}\dket{\alpha}_{\bar x,L},\\(\bst^2)_\alpha=e^{\imth\Phi_{\bar x}}e^{\pi\imth Q_{\alpha,\bar x}}=\pm1.
\eea
where $\Phi_{\bar x}$ is a phase factor independent of Schmidt eigenstates $\dket{\alpha}_{\bar x,L}$. Now let's consider the ground state
\bea
\dket{\Psi_1}\equiv T_x^{-1}\mathcal{F}_y(\pi L_y)\dket{\Psi_0}=e^{\imth\phi_0}\sum_{\alpha}\lambda_{\bar x,\alpha}\dket{\alpha}_{\bar x,L}\dket{\alpha}_{\bar x,R}
\eea
after flux insertion, which has the same Schmidt eigenstates as $\dket{\Psi_0}$ up to a global phase factor. Clearly $\mathcal{F}_y(\pi L_y)\dket{\Psi_0}=T_x\dket{\Psi_1}$ is also a symmetric SRE state with Schmidt decomposition
\bea
&\notag\mathcal{F}_y(\pi L_y)\dket{\Psi_0}=e^{\imth\phi_0}\sum_\alpha\lambda_{\bar x,\alpha}\dket{\alpha}_{\bar x+1,L}\dket{\alpha}_{\bar x+1,R},\\
&\dket{\alpha}_{\bar x+1,L/R}\equiv \hat T_x\dket{\alpha}_{\bar x,L/R}.\label{sdcp:after}
\eea
across an entanglement cut along $\hat y$ direction at $\bar x+1\in(x_0,x_0+1)$. Meanwhile if we choose the same entanglement cut at $\bar x+1$, $\dket{\Psi_0}$ can be Schmidt decomposed into
\bea
\dket{\Psi_0}=\sum_{\beta}s_{\bar x+1,\beta}\dket{\beta}_{\bar x+1,L}\dket{\beta}_{\bar x+1,R}
\eea
where the Schmidt eigenstates of $\dket{\Psi_0}$ at the two different cuts are related by
\bea
\dket{\beta}_{\bar x+1,L}=\sum_{p,\alpha}M^p_{\beta,\alpha}\dket{p}_{x_0}\otimes\dket{\alpha}_{\bar x,L}.
\eea
As shown in Ref.\cite{Watanabe2015} with $\bar\rho_f=$~odd fermions per unit cell, $\dket{\beta}_{\bar x+1,L}$ and $\dket{\alpha}_{\bar x,L}$ have different time reversal representations \ie
\bea
(\bst^2)_{\beta,\bar x+1}=-(\bst^2)_{\alpha,\bar x}.
\eea
Meanwhile from relation (\ref{sdcp:after}) it's straightforward to see that $\dket{\alpha}_{\bar x+1,L}$ and $\dket{\alpha}_{\bar x,L}$ share the same time reversal representation \ie
\bea
(\bst^2)_{\alpha,\bar x+1}=(\bst^2)_{\alpha,\bar x}.
\eea
As a result, the Schmidt eigenstates of many-body ground states before and after flux insertion \ie $\dket{\Psi_0}$ and $\mathcal{F}_y(\pi L_y)\dket{\Psi_0}$ across the same cut at $\bar x+1$ have different time reversal representations
\bea\label{sdcp:trs}
(\bst^2)_{\beta,\bar x+1}=-(\bst^2)_{\alpha,\bar x+1}.
\eea
Notice that generically no pumping is induced by adiabatic flux insertion in a trivial Mott insulator that can change the time reversal representation of Schmidt eigenstates, and hence (\ref{sdcp:trs}) is contradictory to a trivial Mott insulating ground state.

On the other hand, as shown in Ref.\cite{Fu2006,Levin2009}, one Kramers doublet is pumped from one end of the cylinder to the other during the adiabatic $\pi$ flux ($L_y=$~odd) insertion $\mathcal{F}_y(\pi L_y)$. This ``$Z_2$ spin pumping''\cite{Fu2006} can change the Schmidt eigenstates from a Kramers singlet into a Kramers doublet, and is consistent with (\ref{sdcp:trs}). Therefore while a trivial Mott insulator is ruled out by our entanglement spectrum argument, a QSH insulator becomes the only possible symmetric SRE ground state compatible with all symmetries at half filling. Therefore we've proved Theorem 1.

\begin{figure}
\includegraphics[width=0.7\columnwidth]{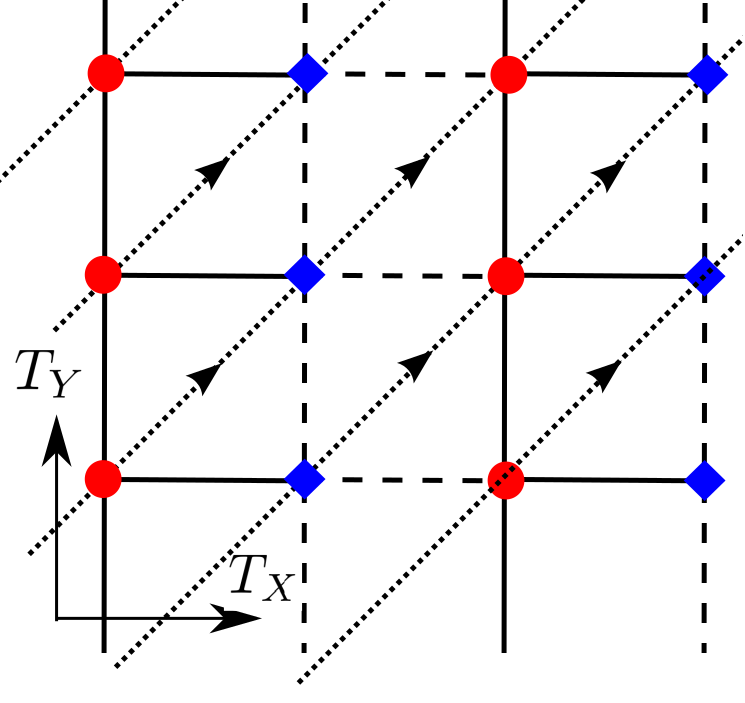}
\caption{(color online) Illustration of $\pi$ flux model (\ref{square model}) of a QSH insulator on square lattice. Solid (dashed) lines denote positive (negative) real hoppings between nearest neighbors (NNs), while dotted lines denote imaginary hoppings \ie spin-orbit couplings between next NNs.}
\label{fig:square pi flux}
\end{figure}

\begin{figure}
\includegraphics[width=0.9\columnwidth]{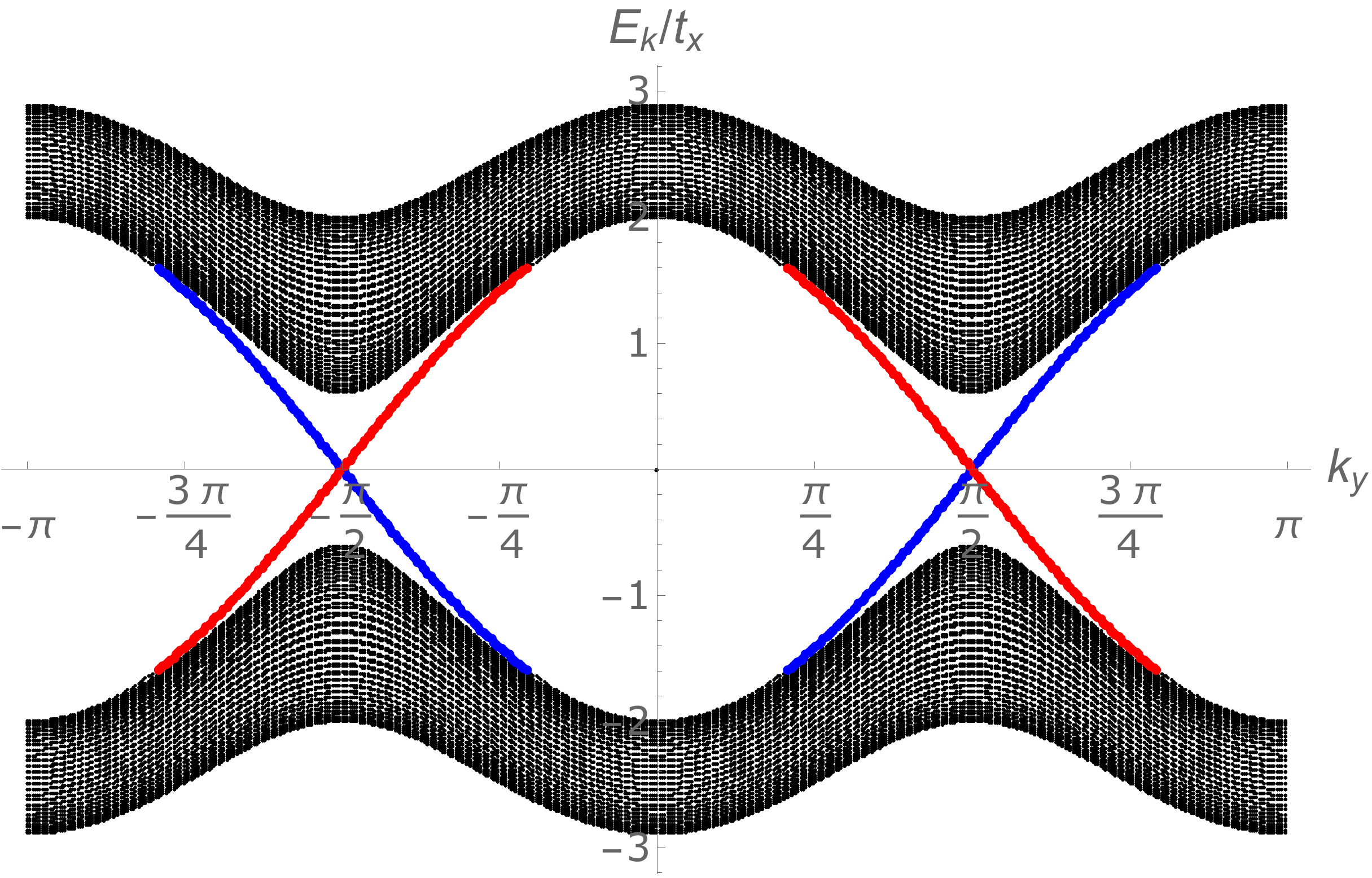}
\caption{(color online) The band structure with open boundary condition (30 magnetic unit cells) in the $x-$direction. Here we set $t_x=t_y=-1.0$, and $t_2=0.3$. The blue line indicates the gapless edge mode localized at the left edge, while the red line indicates the one on the right.}
\label{fig:band_structure_v3}
\end{figure}

\paragraph{$\pi$-flux model on the square lattice---}While Theorem 1 applies to a generic interacting fermion system, here we demonstrate its validity in a non-interacting fermion system for simplicity. In particular we consider the following $\pi$-flux model on square lattice (see FIG. \ref{fig:square pi flux})
\bea
&\notag\hat H_0=\sum_{{\bf r},\sigma}\Big[(-1)^x(t_xf^\dagger_{{\bf r},\sigma}f_{{\bf r}+\hat x,\sigma}+t_yf^\dagger_{{\bf r},\sigma}f_{{\bf r}+\hat y,\sigma})\\
&+\imth\sigma\cdot t_2f^\dagger_{{\bf r},\sigma}f_{{\bf r}+\hat x+\hat y,\sigma}\Big]+h.c.\label{square model}
\eea
It's straightforward to see that it preserves magnetic translation symmetry
\bea
\tilde T_x=T_x(-1)^{\sum_{\bf r}(x+y)\hat n_{\bf r}},~~~\tilde T_y=T_y.
\eea
which differs from (\ref{landau gauge}) merely by a gauge transformation $G_{(x,y)}=(-1)^{\frac{x(x-1)}{2}}$. Choosing doubled magnetic unit cell as $\psi_{(x,y)}\equiv(f_{(2x,y)},f_{(2x+1,y)})^T$, the above Hamiltonian in momentum space writes
\bea
&\notag\hat H_0=\sum_{{\bf k}}\psi^\dagger_{{\bf k}}\Big\{t_x[(1-\cos{k_x})\tau_x+\sin{k_x}\tau_y]+2t_y\cos k_y\tau_z\\
\notag&+t_2\sigma_z[-(\sin k_y+\sin(k_x+k_y))\tau_x\\
&\notag+(\cos k_y-\cos(k_x+k_y))\tau_y]\Big\}\psi_{\bf k}
\eea
where $\vec\tau$ and $\vec\sigma$ are Pauli matrices for sublattice and spin indices respectively. Nearest neighbor real hoppings $t_{x,y}$ leads to two Dirac cones at two ``valleys'' ${\bf k}\simeq\pm(0,\pi/2)$
\bea\label{dirac ham}
\hat H^{Dirac}_{\bf q}=t_xq_x\tau_y-2t_yq_y\tau_z\mu_z-2t_2 \sigma_z\tau_x\mu_z
\eea
while ${\bf q}={\bf k}\mp(0,\pi/2)$ and $\vec\mu$ are Pauli matrices for valley index. Clearly next nearest neighbor spin-orbit couplings ($\imth t_2$ term) open up a QSH mass gap for the Dirac fermions. It's straightforward to figure out the symmetry operations
\bea
R_{\tilde T_y}=\imth\mu_z,~~R_{\tilde T_x}=\imth\tau_y\mu_x,~~R_{\bst}=\imth\sigma_y\mu_x K
\eea
on the Dirac Hamiltonian (\ref{dirac ham}), and therefore the only mass terms allowed by time reversal and magnetic translation symmetries are indeed the 3 QSH masses
\bea
\vec{\bf V}_{QSH}=\vec\sigma\tau_x\mu_z
\eea
as dictated by Theorem \ref{thm}. By imposing open boundary condition, one can see the gapless edge modes in the energy spectrum as shown in FIG. \ref{fig:band_structure_v3}.

\paragraph{Realization in cold atoms---}

Consider a single particle hamiltonian with a time dependent potential and a time dependent magnetic field
\begin{equation}
H = \int {\rm d} {\bf r} \psi^{\dagger}({\bf r} )\left( \frac{ {\bf p}^2}{2M} + V({\bf r}, t) -  B\sigma_{z} {\rm cos}\Omega t \right) \psi({\bf r})
\end{equation}
With a digital mirror device, one can engineer a static potential on a plane of  the form,
\begin{equation}
V({\bf r}) = \sum_{\bf R} U({\bf r}- {\bf R})
\end{equation}
or a shaking potential of the form
\bea
\notag&V({\bf r}) = \sum_{\bf R} U({\bf r}- {\bf R} - {\bf A}({\bf R}, t))\\
&= V({\bf r}) - \sum_{\bf R} {\bf A}({\bf R},t) \cdot
\nabla V({\bf r}- {\bf R}) + ...
\label{Vrt} \eea
where $\{ {\bf R} \}$ form a Bravis lattice.
We shall proceed by only keeping the term of Eq.(\ref{Vrt}), i.e. assuming we are in the perturbative regime. However, our discussions can proceed with an exact calculation. That is more involved and I shall discuss at other time.  The perturbative calculation is valuable because it provides a quick way to
find out whether the hamiltonian of the right symmetry can be constructed.

The hamiltonian is now

\begin{equation}
H(t) = H_{o} + H_{1}(t) + H_{Z}(t)
\end{equation}
\begin{equation}
H_{o}= \int {\rm d} {\bf r} \psi^{\dagger}({\bf r} )\left( \frac{ {\bf p}^2}{2M} + V({\bf r})  \right) \psi({\bf r})
\label{Ho} \end{equation}
\begin{equation}
H_{1}(t) =  - \sum_{\bf R} {\bf A}({\bf R}, t)  \cdot
\int {\rm d}{\bf r}   \psi^{\dagger}({\bf r} )
\left[ \nabla V({\bf r}- {\bf R}) \right]  \psi^{}({\bf r} )
\label{H1} \end{equation}
\begin{equation}
H_{Z}(t) =  - B {\rm cos}\Omega t     \int {\rm d} {\bf r} \psi^{\dagger}({\bf r} ) \sigma_{z}  \psi({\bf r})
\end{equation}

We shall now consider the square lattice case, and take
 the static hamiltonian Eq.(\ref{Ho})  to be   $\hat H_{o} = \hat T +\hat U$
\begin{equation}
\hat T = - t\sum_{{\bf R}, {\bf e}}  \left(  a^{\dagger}_{{\bf R} + {\bf e}, \mu} a^{}_{{\bf R}, \mu}
+ h.c. \right), \,\,\,
\hat U = \sum_{\bf R} {\bf E}\cdot {\bf R}
a^{\dagger}_{{\bf R}, \mu} a^{}_{{\bf R}, \mu}\label{tight binding}
\end{equation}
where ${\bf R}$ forms a square lattice, and ${\bf e} = \hat{\bf x}$  and $\hat{\bf y}$.  Repeated spin indices are being summed over. The term ${\bf E}\cdot {\bf R}$ is a linear static potential.  $a_{{\bf R}, \mu}$ annihilates a particle at site ${\bf R}$ with spin $\mu$.  The corresponding Wannier wavefunction will be denoted as $w_{\bf R}({\bf r})$ and is spin independent.   With the expansion $\psi_{\mu}({\bf r} )= \sum_{\bf R} w_{\bf R}({\bf r})a_{{\bf R}, \mu}$,  Eq.(\ref{H1}) can now be written
\begin{equation}
H_{1}(t) =   \sum_{\bf R,\bf S, \bf S'} {\bf A}({\bf R}, t)  \cdot
{\bf D}_{\bf S, S'}({\bf R}) a^{\dagger}_{{\bf S}, \mu}
 a^{}_{{\bf S'}, \mu}
\end{equation}
\begin{equation}
{\bf D}_{\bf S, S'}({\bf R})  = \int {\rm d} {\bf r} V({\bf r}- {\bf R}) \left[ \nabla (w^{\ast}_{\bf S} ({\bf r}) w^{}_{\bf S'} ({\bf r}) \right]   \label{D} \end{equation}
For a tight binging band, $w_{\bf S}({\bf r})$ can be well approximated by a real Guassian about the lattice site ${\bf S}$. It is then easy to see that the largest term in Eq.(\ref{D}) are those ${\bf S}={\bf R}$ and ${\bf S}={\bf R}+{\bf e}$; or ${\bf S}={\bf R}+{\bf e} $ and ${\bf S}={\bf R}$, and it is easy to show
\begin{equation}
H_{1}(t) = -  \lambda  \sum_{{\bf R}} {\bf A}({\bf R}, t)  \cdot   {\bf e}
\left( a^{\dagger}_{{\bf R}, \mu} a^{}_{{\bf R}+{\bf e} , \mu} + h.c. \right)
\end{equation}
where $\lambda$ is a real number.

Next, we note that $H_{Z}$ can be written as
\bea
\notag&H_{Z} =  - B {\rm cos}\Omega t  \sum_{{\bf R, R'},\mu}  F_{\bf R,R'} a^{\dagger}_{ {\bf R}, \mu}\sigma^{z}_{\mu\mu'} a^{}_{{\bf R'}, \mu'},\\ &F_{\bf R,R'}  =\int {\rm d} {\bf r}  w^{\ast}_{\bf R}({\bf r}) w_{\bf R'}({\bf r})
\eea
The time dependent Schrodinger equation is now
\begin{eqnarray}
\notag&\imth\partial_{t} |\Psi (t)\rangle =  \\
&\left( - t\sum_{{\bf R}, {\bf e}}  a^{\dagger}_{{\bf R} + {\bf e}, \mu} a^{}_{{\bf R}, \mu}  + h.c.    +  \sum_{\bf R} {\bf E}\cdot {\bf R}  a^{\dagger}_{{\bf R} \mu} a^{}_{{\bf R}, \mu}  \right)   |\Psi (t)\rangle
 \nonumber \\
& - \lambda \sum_{\bf R, e} \left(    ( {\bf e}\cdot {\bf A}({\bf R}, t) )  (a^{\dagger}_{{\bf R} + {\bf e}, \mu} a^{}_{{\bf R}, \mu}   + h.c. \right) |\Psi (t)\rangle    \nonumber \\
& - B ({\rm cos}\Omega t )\sum_{\bf R, R'}  F_{\bf R,R'} a^{\dagger}_{ {\bf R}, \mu}\sigma^{z}_{\mu\mu'} a^{}_{{\bf R'}, \mu'} |\Psi (t)\rangle
\end{eqnarray}
Now consider
\begin{equation}
{\bf A}({\bf R}, t) = \sum_{i=1,2} \left( {\bf a}_{i}e^{\imth({\bf Q}_{i}\cdot {\bf R} -\omega_{i} t)} + c.c \right) , \,\,\,\,\,\,\, {\bf a}_{i} \equiv {\bf c}_{i}
+ i {\bf d}_{i}.
\label{A} \end{equation}
where ${\bf c}_{i}$ and ${\bf d}_{i}$ are real vectors.  Eq.(\ref{A}) represents a shaking amplitude that is a linear combination of two oscillations, each one has its own frequencies $\omega_{i}$,   wavevector ${\bf q}_{i}$, and complex poliarization ${\bf a}_{i}$. As we shall, each oscillation ${\bf a}_{i}$ is to generate the desire hoppings in the $x$ and $y$ direction.

Now we perform a unitary transformation
\begin{equation}
|\Psi(t)\rangle \equiv e^{ -  i \hat{U} t}|\Phi(t)\rangle
\end{equation}
where $\hat U$ is defined in (\ref{tight binding}). We have
\begin{widetext}
\begin{eqnarray}
\notag&i \partial_{t} |\Phi (t)\rangle =  \left( - t\sum_{{\bf R}, \mu}  e^{i {\bf E}\cdot{\bf e} t} a^{\dagger}_{{\bf R} + {\bf e}, \mu} a^{}_{{\bf R}, \mu}  + h.c.     \right)   |\Phi (t)\rangle\\
&- \lambda \sum_{\bf R, e} \sum_{i}
\left[ {\bf e}  \cdot    ({\bf a}_{i}e^{i({\bf Q}_{i}\cdot {\bf R} -\omega_{i} t)} + c.c.)   \right]
 \left[  e^{i {\bf E}\cdot{\bf e} t}
 a^{\dagger}_{ {\bf R} + {\bf e}, \mu} a^{}_{ {\bf R}, \mu}   + h.c.  \right]  |\Phi (t)\rangle
 \nonumber \\
& - B ({\rm cos}\Omega t )\sum_{\bf R, R'}  F_{\bf R,R'} e^{i {\bf E}\cdot ({\bf R}-{\bf R'})t }
 a^{\dagger}_{ {\bf R}, \mu}\sigma^{z}_{\mu\mu'} a^{}_{{\bf R'}, \mu'} |\Phi (t)\rangle
 \equiv \left[ (1^{st}) + (2^{nd}) + (3^{rd})\right] |\Phi(t)\rangle     \equiv {\cal H}  |\Phi(t)\rangle
\label{trans} \end{eqnarray}
\end{widetext}

Consider now
 \begin{equation}
\omega_{1} = E_x, \,\,\,\,\, {\bf a}_{1} =  c_{1} \hat{\bf x}, \,\,\,\,\,\, {\bf Q}_{1} = q\hat{\bf x},
\end{equation}
\begin{equation}
\omega_{2} = E_y, \,\,\,\,\, {\bf a}_{2} = d_{1} \hat{\bf y}, \,\,\,\,\,\, {\bf Q}_{2} = q\hat{\bf x}.
\end{equation}
\begin{equation}
\Omega = E_{x} + E_y,
\end{equation}

The hamiltonian ${\cal H}$ in Eq.(\ref{trans}) then consists of both time independent term
${\cal K}$ and time dependent terms ${\cal K}_{1}$.  The time independent terms are
\bea
\notag&{\cal K} =  \sum_{\bf R} \Big[  -  \lambda   e^{\imth qR_{x}}( c_{1} a^{\dagger}_{{\bf R}+\hat{\bf x}, \mu} a^{}_{{\bf R}, \mu} +d_{1}  a^{\dagger}_{{\bf R}+\hat{\bf y}, \mu} a^{}_{\bf R, \mu}) \\
&- BF a^{\dagger}_{{\bf R} + \hat{\bf x} + \hat{\bf y}, \mu}\sigma^{z}_{\mu\mu'} a^{}_{\bf R, \mu'}   + h.c.\Big]\label{K}
\eea
where $F=F_{{\bf R}, {\bf R}+ \hat{\bf x} + \hat{\bf y}}$.
The time dependent terms ${\cal K}_{1}$ all contains oscillating factors of the form $e^{i\omega t}$.
Since there are no resonances that are caused by these terms, they will be averaged to zero.
So we can focus on ${\cal K}$.
It's straightforward to see that (\ref{K}) is the desired Hamiltonian (\ref{square model}) if $q=\pi$, which makes the factor $e^{\imth qR_{x}} = (-1)^{R_x}$.

\paragraph{Conclusions---}

In this letter, we established a Lieb-Schultz-Mattis theorem for QSHE at half filling, enforced by magnetic translation symmetry with $\pi$-flux per unit cell. This theorem sheds new light in realizing QSHE in strongly interacting systems. We demonstrate the theorem by a square-lattice $\pi$-flux model, which we show can be realized in cold atoms in a shaking lattice under a time-dependent Zeeman field.

\acknowledgments
YML thanks Ying Ran and Masaki Oshikawa for related collaborations, and KITP ``topoquant16'' program for hospitality
where part of this work was finished. This work is supported by xxxxxxxx (JW, TLH), Startup Funds at OSU (YML), and in part by NSF under Grant No. NSF PHY11-25915 (YML).

\bibliographystyle{apsrev4-1}
\bibliography{bibs}

\appendix

\end{document}